\documentclass[conference]{IEEEtran} 

\newcommand{\cmmnt}[1]{}  

\IEEEoverridecommandlockouts 

\usepackage{amsmath} 

\usepackage{comment}
\usepackage{blindtext}
\usepackage{dirtytalk}
\usepackage{array}
\usepackage{authblk}
\usepackage{url}
\usepackage[noadjust]{cite}
\usepackage{graphics}
\usepackage{graphicx}
\usepackage{epsfig}
\usepackage{amssymb}
\usepackage{amsmath}
\usepackage{breqn}
\usepackage[ruled,vlined]{algorithm2e}

\usepackage{mathtools}
\usepackage{lscape}
\usepackage{pdflscape}
\usepackage{longtable}
\usepackage{makecell}
\usepackage{lettrine}
\usepackage{caption}
\usepackage{supertabular,booktabs}

\usepackage{amsmath}
\usepackage{graphicx}
\usepackage[colorlinks=true, allcolors=blue]{hyperref}
\usepackage{amssymb}
\usepackage[table,xcdraw]{xcolor}
\usepackage{tabularx} 
\usepackage{subfigure}

\usepackage{color,soul}
\usepackage{multirow}

\newcolumntype{M}{>{$\displaystyle}c<{$}} 
\newcommand\AddLabel[1]{\refstepcounter{equation}(\theequation)\label{#1}}
\newcolumntype{L}{>{\collectcell\AddLabel}r<{\endcollectcell}}

\title{\LARGE 
Enhancing Manufacturing Training Through VR Simulations
}

\begin{document}


\author{Vladislav Li\IEEEauthorrefmark{1}, Ilias Siniosoglou\IEEEauthorrefmark{2}\IEEEauthorrefmark{3},Panagiotis Sarigiannidis\IEEEauthorrefmark{2}\IEEEauthorrefmark{3} and Vasileios Argyriou\IEEEauthorrefmark{1}

\thanks{\IEEEauthorrefmark{1} V. Li and V. Argyriou are with the Department of Networks and Digital Media, Kingston University, Kingston upon Thames, United Kingdom - \texttt{E-Mail: \{v.li, vasileios.argyriou\}@kingston.ac.uk}}

\thanks{\IEEEauthorrefmark{2} I. Siniosoglou and P.Sarigiannidis are with the Department of Electrical and Computer Engineering, University of Western Macedonia, Kozani, Greece - \texttt{E-Mail: \{isiosoglou,psarigiannidis\}@uowm.gr}}

\thanks{\IEEEauthorrefmark{3} I. Siniosoglou and P. Sarigiannidis are with MetaMind Innovations P.C., Kozani, Greece - \texttt{E-Mail: \{isiosoglou,psarigiannidis\}@metamind.gr}}

}

\maketitle
\thispagestyle{empty}
\pagestyle{empty}

\begin{abstract}
In contemporary training for industrial manufacturing, reconciling theoretical knowledge with practical experience continues to be a significant difficulty. As companies transition to more intricate and technology-oriented settings, conventional training methods frequently inadequately equip workers with essential practical skills while maintaining safety and efficiency. Virtual Reality has emerged as a transformational instrument to tackle this issue by providing immersive, interactive, and risk-free teaching experiences. Through the simulation of authentic industrial environments, virtual reality facilitates the acquisition of vital skills for trainees within a regulated and stimulating context, therefore mitigating the hazards linked to experiential learning in the workplace. This paper presents a sophisticated VR-based industrial training architecture aimed at improving learning efficacy via high-fidelity simulations, dynamic and context-sensitive scenarios, and adaptive feedback systems. The suggested system incorporates intuitive gesture-based controls, reducing the learning curve for users across all skill levels. A new scoring metric, namely, VR Training Scenario Score (VRTSS), is used to assess trainee performance dynamically, guaranteeing ongoing engagement and incentive. The experimental assessment of the system reveals promising outcomes, with significant enhancements in information retention, task execution precision, and overall training efficacy. The results highlight the capability of VR as a crucial instrument in industrial training, providing a scalable, interactive, and efficient substitute for conventional learning methods.

\end{abstract}

\begin{IEEEkeywords}
Virtual Reality Training, User Engagement, Immersive Learning, Interactive Simulations, Health and Safety Education, Task Performance, Training Effectiveness, Industrial Manufacturing
\end{IEEEkeywords}

\section{Introduction}
\label{Introduction}

In the modern industrial era, with the transition to Industry 5.0 and its emphasis on human-centric, sustainable operations, the organization of manufacturing environments is undergoing significant evolution. This transformation spans from the physical layer of equipment and sensor deployments to the digital layer of data processing and decision-making systems \cite{machines12080528, li2025enhancing}. One major driver of this shift is the integration of Virtual Reality (VR) technologies, which streamline training, design, and operational workflows. The high throughput of data generated in real-time manufacturing processes underscores the need for immersive solutions \cite{hell2018machine}. VR’s ability to provide intuitive, context-aware simulations ensures a seamless link between virtual prototypes and real-world implementation, offering industries tools to bridge gaps in worker training, process optimization, and operational safety \cite{Porcino6_37}.

The integration of VR technology into manufacturing training represents a transformative leap forward in bridging theoretical knowledge with practical application \cite{Rafa2024}. This study explores how VR simulations can enhance training effectiveness, accessibility, and user engagement in manufacturing environments. By offering immersive, context-aware scenarios tailored to industry-specific needs, VR training modules empower employees to gain hands-on experience in a safe and controlled setting \cite{montenegro2017cognitive,li2024closer}.

Despite the promise of VR in manufacturing training, several challenges \cite{Porcino6_37} have hindered its widespread adoption. One major difficulty lies in ensuring the scalability of VR solutions across diverse manufacturing environments. Manufacturing operations often vary widely in complexity, requiring adaptable training modules that reflect specific workflows, equipment configurations, and safety protocols. Developing a system capable of accommodating such variability, while maintaining the realism and effectiveness of the training experience, demands significant technical and design expertise \cite{Lupinetti2019}. Furthermore, achieving high-quality, interactive XR simulations involves overcoming hardware limitations, such as ensuring compatibility with a wide range of XR devices and optimizing system performance to prevent latency or discomfort during use \cite{10869099,nagy2025}.

Another significant challenge is ensuring that VR systems remain user-friendly for a workforce that may have limited familiarity with advanced digital technologies \cite{li2023modular}. Many trainees, particularly those with minimal exposure to VR systems, require orientation sessions to navigate the immersive environment effectively. The steep learning curve for some users can detract from the training experience, necessitating the design of interfaces that are intuitive and accessible across varying skill levels. Additionally, the reliance on real-time data processing for personalized feedback introduces challenges related to data integrity and security. Manufacturing environments often deal with sensitive operational data, requiring robust encryption and access controls to safeguard proprietary information. Addressing these challenges is critical for unlocking the full potential of VR as a tool for improving manufacturing training outcomes. To this end, this paper focuses on overcoming the aforementioned challenges, effectively aiming to create an advanced and high fidelity VR training ecosystem for the high risk - high pace industrial environment. Specifically, the contributions of this paper can be summarized as follows:

\begin{itemize}
\item An advanced scheme and methodology for industrial manufacturing VR training with context-aware scenarios that mimic real-world environments and enable trainees to better comprehend learning materials and situations.
\item A high fidelity and realistic simulation environment for enhancing manufacturing training situational awareness with adaptive feedback mechanisms.
\item A methodology for dynamic scenario adaptation based on user feedback and actions.
\item Natural gesture-based controls and minimal learning curve interfaces ensure seamless navigation for users of all skill levels, complementing the immersive experience.
\item A Novel scoring mechanism integrated into interactive scenarios to evaluate and maintain user motivation.
\end{itemize}

The rest of this paper is organized as follows: the related work is discussed in Section \ref{Literature Review}, followed by an overview of the methodology in Section \ref{Methodology}. Section \ref{Evaluation} provides a comprehensive analysis of the available data, as well as a series of quantitative results. Section \ref{Conclusions} offers concluding remarks.


\section{Literature Review}
\label{Literature Review}

Due to the powerful nature of VR in training employees, the technology has seen a big rise in implementations, especially in the industrial domain. This is powered by the fact that safety is essential in such domains but so is the need to purposefully train and evaluate employees before actually giving aces to dangerous and of course delicate industrial equipment. To this end, a lot of related work has been seen in this field.

For example, in \cite{lie2023trainingopenendeddrillingvirtual} the authors propose a VR training system to teach open-ended psychomotor skills for drilling with a 3-axis milling machine. The simulation includes modules for safety, tutorials, open-ended practice, and evaluation against engineering standards, enabling multiple pathways for task completion. A user study with engineering students showed higher task success rates, fewer mistakes, and faster completion times compared to traditional methods. The study also highlights the cognitive, psychomotor, and affective benefits of open-ended VR training, offering a novel approach to enhancing industrial skill acquisition through flexible, exploratory learning.Despite its effectiveness, the methodology may face challenges in industrial settings, where the diversity of equipment and environmental variability could hinder its generalizability. 

Likewise, in \cite{noghabaei2019virtualmanipulationimmersivevirtual}, the authors propose a VR platform to enhance the assembly and inspection of industrial components using immersive virtual environments (IVEs). The system integrates BIM models and 3D laser-scanned data, enabling users to manipulate and evaluate parts for discrepancies. Leveraging a motion-tracked VR setup, the platform highlights defective components and ensures quality control by snapping components into place if within defined tolerances. This approach facilitates training, reduces on-site errors, and streamlines quality assurance processes for complex assemblies, offering significant time and cost savings. While the proposed VR platform is effective for training and quality control, it may face challenges in scaling to larger, more intricate industrial settings with highly complex assemblies. The reliance on 3D scanning accuracy and potential noise in point cloud data could limit its precision in detecting discrepancies. Similarly, in \cite{Hanifati10791222} the authors present a VR-based Operator Training Simulator (VR OTS) designed to simulate normal, startup, and shutdown conditions in manufacturing environments. Users can train as field or control panel operators, engaging in immersive scenarios that replicate real-world tasks. Utilizing the Meta Oculus Quest 2, the system integrates action indicators and interactive environments to enhance operational efficiency and safety.

Another interesting implementation can be seen in \cite{Rafa2024}, where the authors present a VR platform for additive manufacturing (AM) education, simulating a 3D printing lab with hands-on tasks like parameter selection, safety compliance, and cost estimation. The study shows improved student attention, accuracy, and satisfaction, though challenges include immersive stimuli and limited feedback during decision-making. Similarly, \cite{Nasri10765352} introduces a Virtual Reality Training Apprenticeship (VRTA) for the cold spray process, featuring six modules, including powder feeder assembly. The VRTA emphasizes realism and interactive instructions, effectively transferring skills to real-world tasks. However, limitations like lack of physical tool realism, non-intuitive interactions, and a small sample size hinder wider applicability.

\section{Methodology}
\label{Methodology}
\begin{figure}[h]
 \centering
  \includegraphics[scale=0.55]{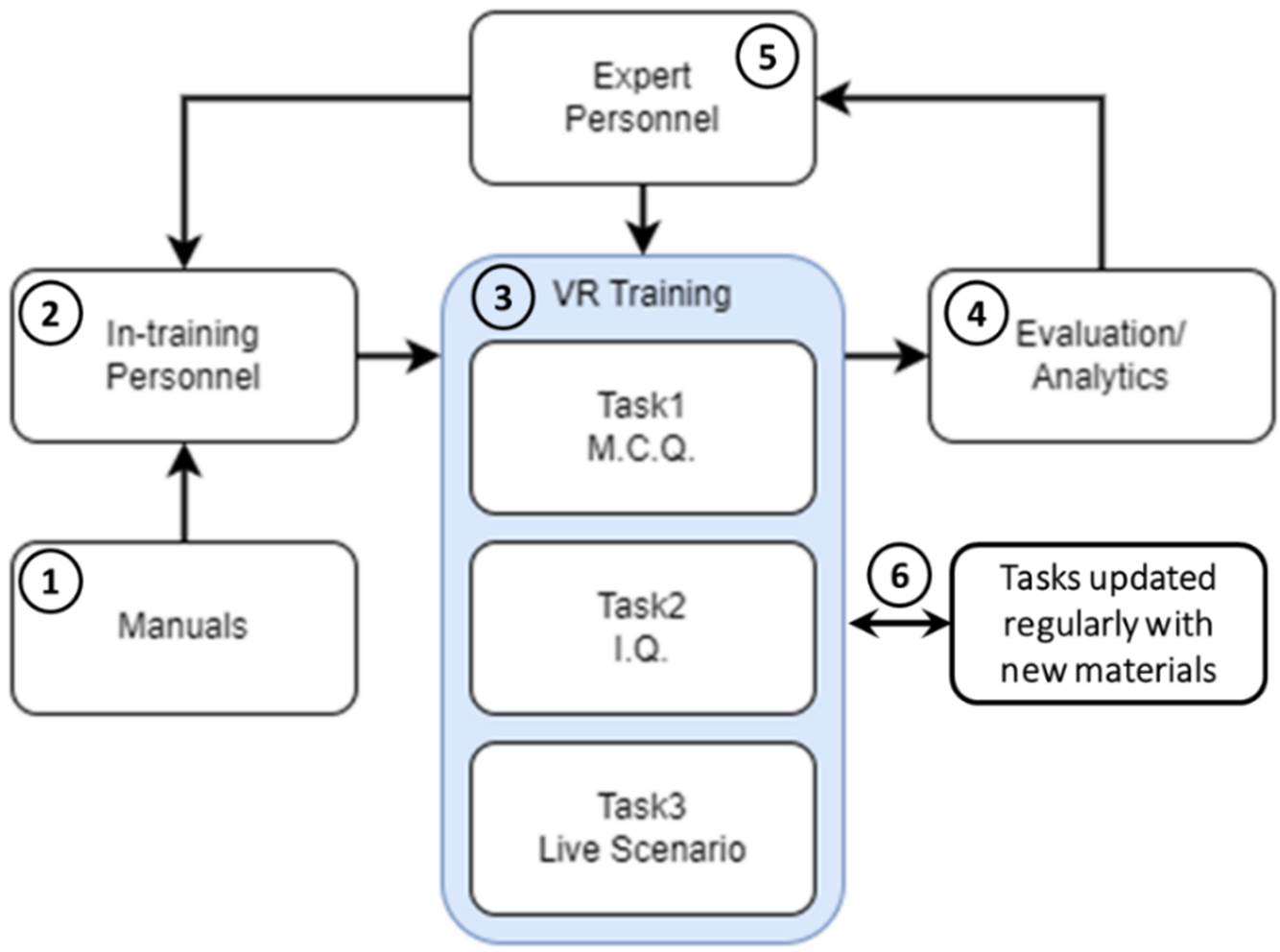}
 \caption{Live Scenario Scene Assets: Real-time dynamic scenarios simulated complex industrial decision-making, challenging trainees to handle unexpected events like equipment failures or safety hazards while adapting to their responses. MCQ is for Multiple-Choice Questions, and IQ is for Interactive Questions}
 \label{fig:methodology}
\end{figure}

The proposed methodology integrates VR technologies to enhance training in industrial environments, focusing on immersive, context-aware learning experiences. The framework is designed to simulate real-world scenarios, enabling users to practice critical protocols, such as emergency responses, in a safe, controlled environment. Gesture-based controls and user-friendly interfaces are employed to ensure accessibility for users of varying technical proficiencies. This section details the development of VR modules, including scenario design, interaction mechanisms, and evaluation strategies. Emphasis is placed on aligning the simulations with industry-specific needs to maximize relevance, while also addressing metrics such as training effectiveness, user engagement, and knowledge retention. An outline of the proposed methodology is depicted in Figure~\ref{fig:methodology}, also showing the steps followed in sequence.

\subsection{3D Assets for Advanced Realism}

The initial development of the VR training environment relied on developing the assets to represent the manufacturing setting from scratch with the use of 3D primitives such as cubes and spheres within Unity. This decision was made with the goal to speed up the initial prototyping phase and start experimenting with the interactive features introduced with the VR technology. Furthermore, it was largely driven by resource constraints as it was difficult to find highly specific 3D assets for the industrial domain in order to create detailed, custom-designed, and realistic industrial equipment to provide the necessary realism to each scenario. To this end, Unity's built-in primitives were adapted to simulate industrial equipment and tools in a manner that allowed for effective interaction with the user.

To enhance the simulation's realism and usability, third-party assets were integrated into the environment wherever possible, originating from available designs of industrial equipment and specifications based on the developed scenarios. However, the availability of free, high-quality industrial assets online proved to be significantly limited. The scarcity of assets tailored specifically for manufacturing environments necessitated the use of generic or simplified alternatives. For instance, rather than incorporating fully detailed industrial machinery, simplified custom-developed assets were used. These simplified assets provided a functional representation of the training environment without detracting from the core objective of enabling user interaction and immersive learning.

At a later stage, hyper-realistic 3D assets were introduced to closely, if not entirely, resemble real-world industrial equipment. These assets significantly enhanced the authenticity of the training environment, allowing users to engage with highly detailed and accurate representations of machinery and tools. The improved level of detail contributed to a more immersive and effective training experience, bridging the gap between virtual simulation and real-world application.

\begin{figure}[h]
 \centering
  \subfigure[POV 1]{\includegraphics[scale=0.3]{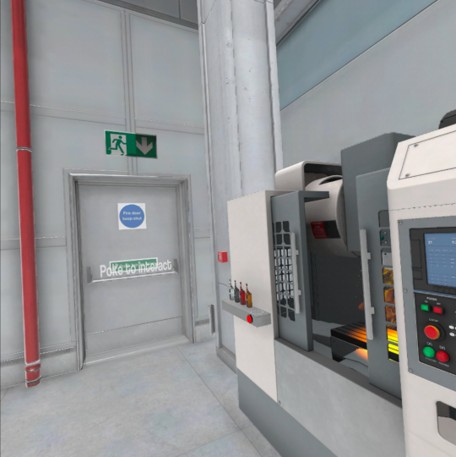}}
  \subfigure[POV 2]{\includegraphics[scale=0.3]{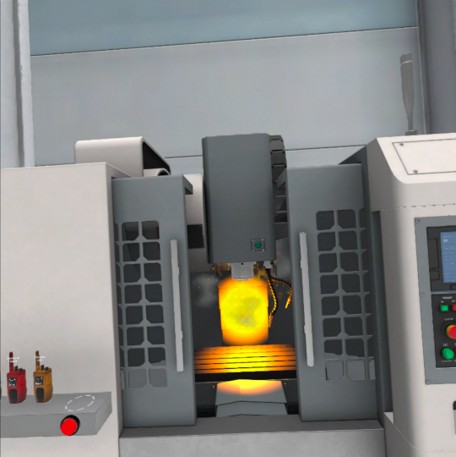}}
 \caption{Live Scenario Scene at different points of view (POVs).}
 \label{fig:3d-assets}
\end{figure}

Despite these limitations, the chosen approach successfully conveyed the concept of interacting with objects and navigating industrial workflows in a VR setting. As can be seen in Figure~\ref{fig:3d-assets}, realistic objects, such as basic levers and industrial boxes, were arranged to create a plausible training scenario. By focusing on functional representation rather than high-fidelity visuals, it was ensured that users could engage with the training scenarios meaningfully.


\subsection{Integration of Interactivity}

 To enhance user experience and convey the needed knowledge to trainees, interactive elements were also incorporated into the VR training modules enhancing engagement and providing an immersive learning experience. The following components formed the core of this interactive framework:

\subsubsection{Questionnaires}
Interactive questionnaires were embedded within the VR modules to evaluate trainee knowledge and understanding of industrial protocols. These quizzes were strategically placed at key points during the training to reinforce learning objectives and assess retention. The questions were designed to address theoretical knowledge, practical application, and problem-solving skills, ensuring comprehensive evaluation. The VR environment enabled a unique approach to questionnaires by presenting them in a dynamic and immersive manner. For instance, users could interact with virtual panels or screens to select answers, making the assessment process engaging and intuitive. This immediate feedback mechanism also allowed trainees to correct misconceptions in real time, fostering a deeper understanding of the material. The integration of questionnaires not only measured progress but also motivated trainees to actively engage with the content throughout the training session.

\subsubsection{Interactable Objects}
Virtual objects, modelled after tools and equipment found in industrial settings, were integrated into the VR modules to provide a hands-on learning experience. These objects allowed trainees to practice tasks such as operating machinery, handling tools, or assembling components. The design prioritized realism and functionality, enabling trainees to interact with the virtual objects as they would in a real-world setting. For example, trainees could grasp, rotate, and position objects using natural hand gestures, facilitated by VR controllers and tracking systems. This immersive interaction helped bridge the gap between theoretical knowledge and practical application. Furthermore, these interactable objects (Figure~\ref{fig:MCQ Scene}) were incorporated into training scenarios to simulate workflows and procedures, ensuring that trainees developed both technical skills and situational awareness. The inclusion of such realistic and functional virtual tools significantly enhanced the overall effectiveness of the training program by providing a tangible and engaging way to practice critical tasks. These developed modules facilitated intuitive user-object interaction, replicating real-world handling of industrial tools and equipment in a virtual setting. By integrating these interactive elements, the VR environment offered an engaging and immersive platform for industrial training.

\subsubsection{Live Scenarios}
\begin{figure*}[ht]
\centerline{\includegraphics[scale=0.095]{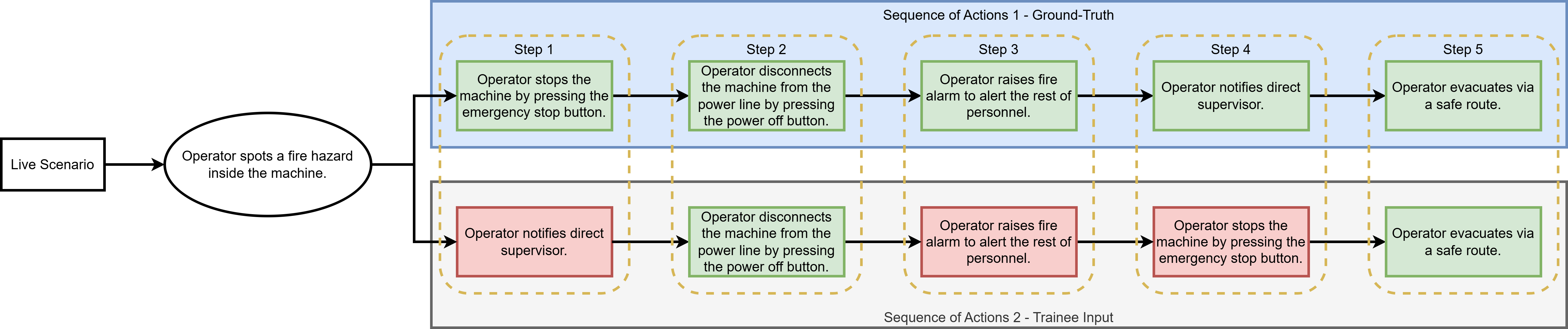}}
\caption{An example sequence of actions for live scenarios in the VR training application. For example, trainees had to identify simulated equipment failures and perform timely repairs, while their actions were assessed for accuracy and sequence.}
\label{fig:vr_training_diagram}
\end{figure*}

Real-time dynamic scenarios were designed to simulate complex, decision-making situations in the industrial context. These scenarios challenged trainees to respond effectively to unexpected events, such as equipment malfunctions or safety hazards, while dynamically adapting the scenario to the trainees responses. By replicating real-world conditions, live scenarios encouraged critical thinking and problem-solving. The VR environment provided a safe space for trainees to experiment with different strategies without the risk of real-world consequences. For example, trainees might need to identify a simulated equipment failure and execute the appropriate repair procedure under time constraints, while evaluating the accuracy and timeliness of trainee feedback through order of actions (see Figure~\ref{fig:vr_training_diagram}) . This interactive and immersive experience enables trainees to build confidence and readiness for actual workplace situations. Additionally, the adaptive nature of the scenarios allowed for difficulty levels to be adjusted based on trainee performance, ensuring personalized and progressive learning outcomes.

\subsection{High Fidelity Interaction Methodology}

To enable a seamless and immersive VR training experience, a sophisticated interaction methodology was developed. This methodology centered around the design of three interactive modules: Multiple-Choice Questions (MCQs), Interactive Questions (IQs), and Live Scenarios. The system aimed to integrate dynamic question-and-answer functionalities, interactive object handling, and adaptive real-time scenarios, enhancing engagement and learning outcomes for trainees.

\begin{figure}[htp]
\centering
  \subfigure[POV 1]{\includegraphics[scale=0.305]{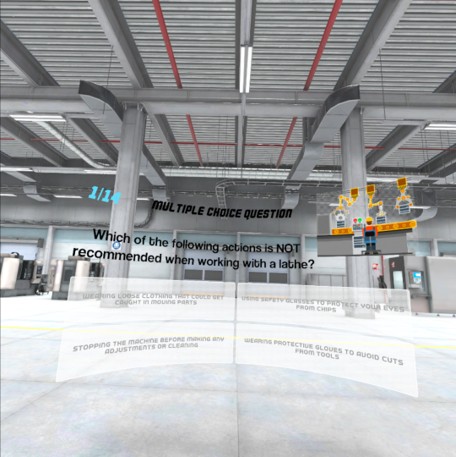}}
  \subfigure[POV 2]{\includegraphics[scale=0.305]{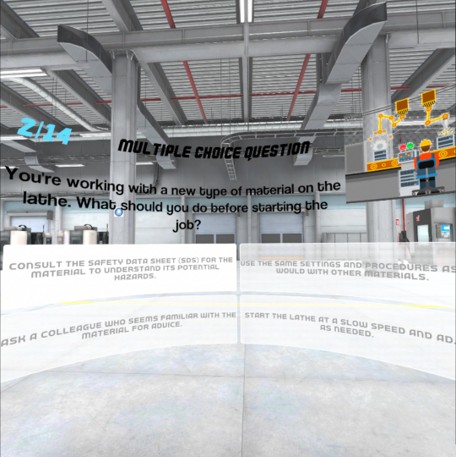}}
 \caption{MCQ Scene at different points of view (POVs)}
 \label{fig:MCQ Scene} 
\end{figure}

\subsubsection{Multiple-Choice Questions (MCQs)}
The MCQ module was designed to facilitate knowledge assessment in an engaging and intuitive manner. The system presented trainees with a series of MCQs that included visual assets and contextual elements. To achieve this, an interactive user interface (UI) was implemented, as can be seen in Figure~\ref{fig:MCQ Scene}, dynamically synchronized with question and answer databases. The module utilized data structures to store question metadata, such as text, asset details, and correctness flags. Real-time updates ensured seamless retrieval of questions and answers, while scoring and progress metrics were logged through a metrics manager. This design not only evaluated user understanding but also provided instant feedback to reinforce learning.

\subsubsection{Interactive Questions (IQs)}
\begin{figure}[htp]
\centering
  \subfigure[POV 1]{\includegraphics[scale=0.308]{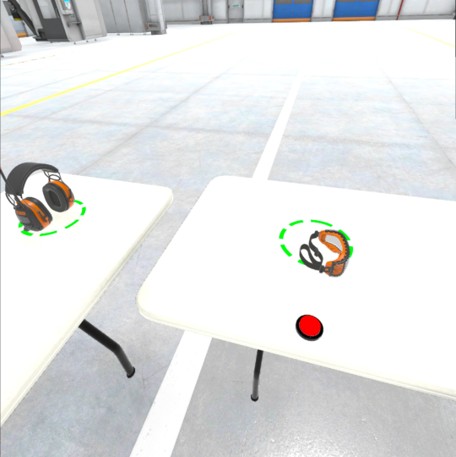}}
  \subfigure[POV 2]{\includegraphics[scale=0.308]{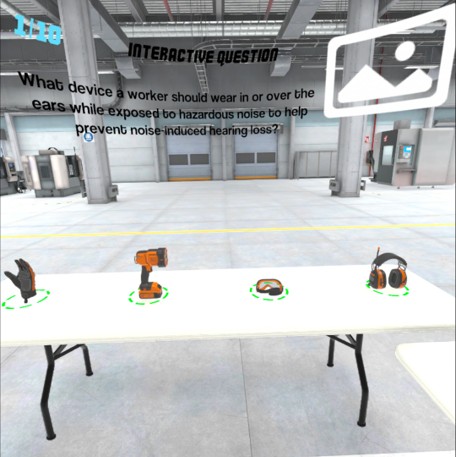}}
 \caption{IQ Scene at different points of view (POVs)}
 \label{fig:IQ Scene} 
\end{figure}

The IQ module expanded the training framework by integrating visual and spatial interaction elements, as seen in Figure~\ref{fig:IQ Scene}. These questions required users to engage with the environment, such as selecting or manipulating virtual objects. A dynamic asset management system allowed the loading of question-related visuals and interaction points. This enhanced realism and ensured that trainees could physically interact with training elements, improving skill acquisition. Additionally, the system supported live scenarios, dynamically adapting IQ content to match real-time situations, further enhancing engagement.

\subsubsection{Live Scenarios}
The live scenario module introduced real-time decision-making tasks to simulate complex industrial situations. Scenarios were structured to challenge users by requiring them to follow specific action sequences or resolve operational challenges. The system tracked user actions, recorded metrics, and evaluated performance against predefined benchmarks. By integrating visual instructions, interactive components, and dynamic feedback, this module replicated real-world scenarios while maintaining a safe and controlled environment.

\subsubsection{System Architecture and Implementation}
The development of these modules was supported by C\# scripting in Unity, leveraging modular and reusable code structures. Two components, the MCQ and IQ Manager modules were developed to enabled managing user interactions and visual asset synchronization. API-based asynchronous calls enabled efficient data retrieval and asset integration, ensuring a responsive and lag-free user experience. Visual elements, such as asset bundles and answer toggles, were seamlessly incorporated into the VR interface, offering an immersive and engaging training environment.

\subsection{Active Metric Acquisition \& Dynamic Scenario Adaptation}

A proposed methodology for active metric acquisition and dynamic scenario adaptation was developed to enable real-time tracking, evaluation, and adaptation within the VR training environment. This system leverages advanced metric collection techniques to monitor user performance and dynamically modify scenarios to enhance engagement and learning outcomes.

\subsubsection{Active Metric Acquisition}

The core of the system’s data collection is managed by a metrics manager component, which tracks and records user interactions, task performance, and engagement metrics. A unique session identifier, generated during initialization, ensures data consistency for each user session. Metrics are categorized into two primary types: performance metrics and user experience (UX) metrics.

Performance metrics include timestamps, task completion data, success rates, and scores, which are logged in real time. These metrics provide insights into trainee progress and proficiency. For instance, timestamps track the duration of specific tasks, while success flags and scores quantify user performance in live scenarios. Data is serialized into a standardized format for local storage and can also be uploaded to an online server for centralized analysis.

UX metrics focus on user engagement and interface usability. These include session-specific metadata such as task names and user interaction data, which are used to evaluate and refine the training experience. Both types of metrics are uploaded to online databases for further processing, allowing for comprehensive reporting and the identification of patterns that inform future development.

\subsubsection{Dynamic Scenario Adaptation}

Dynamic scenario adaptation is a key feature enabled by real-time metric acquisition. During live scenarios, the system monitors user behaviour and compares it against predefined benchmarks, such as the correct sequence of actions. A metrics manager uses this data to dynamically adjust the complexity or structure of scenarios. For example, if a user demonstrates difficulty completing tasks, the system may reduce scenario complexity or provide additional guidance. Conversely, if a user performs exceptionally well, more challenging tasks can be introduced to maintain engagement and encourage growth.

This adaptability is supported by the seamless integration of data retrieval and evaluation systems within the VR environment. The metrics manager processes metrics such as task completion order, timestamps, and success rates to assess performance in real time. The system then uses this information to modify scenario parameters dynamically, ensuring an optimized and personalized training experience for each user. The collected metrics can be broadly categorized into two types: performance metrics and user experience (UX) metrics. Performance metrics provide quantitative data on user proficiency, while UX metrics focus on evaluating the quality of the user interface and interaction design. The adopted metrics are analysed below:  

\subsubsection{Performance Metrics}  

\begin{enumerate}  
    \item \textbf{Task Completion Data}: This metric records whether specific tasks were completed successfully during the training session, providing a binary outcome for task progress \cite{Nath2022}:  
    \begin{equation}
        C = 
        \begin{cases} 
            1 & \text{if task completed successfully} \\
            0 & \text{if task failed or incomplete} 
        \end{cases}
    \end{equation}  
    The sum of all completed tasks across sessions gives an overall completion rate for the training.  

    \item \textbf{Timestamps}: Measures the duration \(T_i\) taken by a user to complete a specific task or scenario. The average task completion time for \(n\) tasks is calculated as \cite{mti6010006}:  
    \begin{equation}
        \text{Average Task Time} = \frac{\sum_{i=1}^{n} T_i}{n}
    \end{equation}  

    \item \textbf{Success Rates}: Tracks the proportion of successfully completed tasks relative to the total number of tasks attempted \cite{joyner2021comparison}:  
    \begin{equation}
        \text{Success Rate} = \frac{\text{Number of Successful Tasks}}{\text{Total Tasks Attempted}} \times 100\%
    \end{equation}  

    \item \textbf{User Performance Score}: Quantifies user performance by assigning weighted scores \(S_i\) to completed tasks. For live scenarios, scores may be normalized using complexity factors \(W_i\) \cite{Vaughan_2020}:  
    \begin{equation}
        \text{Score} = \sum_{i=1}^{n} S_i \cdot W_i
    \end{equation}  

    \item \textbf{Order of Actions}: Logs the sequence of user actions and calculates the similarity between user actions and the correct procedural order using sequence matching techniques, such as \cite{Bueckle_2022}:  
    \begin{equation}
        \text{Order Accuracy} = \frac{\text{Correct Actions in Order}}{\text{Total Expected Actions}} \times 100\%
    \end{equation}  
\end{enumerate}  

\subsubsection{User Experience (UX) Metrics}  

\begin{enumerate}  
    \item \textbf{Engagement Levels}: Assesses user activity and attention by analysing interaction frequency \(F\), calculated as the number of interactions per unit time \cite{Walls2024}:  
    \begin{equation}
        \text{Engagement Frequency} = \frac{\text{Total Interactions}}{\text{Session Duration}}
    \end{equation}  

    \item \textbf{Interaction Data}: Captures details of user actions, such as selecting objects or answering questions, to analyse interface usability and responsiveness. Interaction metrics are aggregated over time to identify usage trends \cite{Tusher2024}.  

    \item \textbf{Session Metadata}: Includes task names and user activities within a session, providing context for performance and engagement data. While no specific formula applies, this data supports qualitative assessments of user experience \cite{murphy2019secondaryinputsmeasuringuser}.  
\end{enumerate}  

These metrics were chosen for their ability to quantitatively and qualitatively assess both the effectiveness of the training system and the user experience. By leveraging these metrics and their derived formulas, the VR framework ensures an adaptive and personalized training environment that meets the diverse needs of its users. In order to quantify better the performance of each trainee, selected performance metrics were utilised to create a unified evaluation metric which is outlined below.

\subsection{VR Training Scenario Score (VRTSS)}

The live scenario module integrates a structured scoring methodology to assess trainee performance in a dynamic and objective manner. This approach ensures that both situational correctness and sequence order are evaluated to provide a comprehensive assessment of trainee proficiency.

To better assess user performance within VR training scenarios, this paper introduces a novel metric known as the VRTSS. This metric is designed to evaluate both procedural accuracy and the correct sequencing of actions, offering a robust measure of learning effectiveness.

The VRTSS formula is defined as:

\begin{equation}
VRTSS = 0.3X + 0.2Y + \sqrt{0.25XY}
\end{equation}

Here, $X$ denotes the correctness of the order of situations, measured as the proportion of correctly sequenced steps. $Y$ denotes the correctness of individual actions, capturing the raw accuracy of trainee decisions. The square root term introduces a geometric mean component to balance the impact of both correctness measures. To evaluate the correctness of the order of situations ($X$), a sequential matching algorithm compares the user’s execution order against the predefined correct sequence. Mathematically, this is expressed as:

\begin{equation}
X = \frac{1}{N} \sum_{i=1}^{N} \delta(S_i, P_i)
\end{equation}

Here, $N$ is the total number of situations in the scenario $S_i$ represents the correct order of the $i^{th}$ situation and $P_i$ denotes the order in which the trainee executed the $i^{th}$ situation.

$\delta(S_i, P_i)$ is an indicator function:

\begin{equation}
\delta(S_i, P_i) =
\begin{cases}
1, & \text{if } S_i = P_i \\
0, & \text{otherwise}
\end{cases}
\end{equation}

This function ensures that only correctly placed situations contribute to the correctness score. The total count of correctly placed situations is then normalized over $N$ to produce a proportion-based metric.

To calculate $X$, the input from the trainee is compared with the ground-truth, as demonstrated in the Figure~\ref{fig:vr_training_diagram}. For instance, if the input from the trainee differs from the ground-truth then the trainee failed to perform an action when it is needed. Likewise, if the trainee’s input matches the ground-truth then the action was performed right on time, i.e., in the right sequence. The VR training application keeps track of trainee’s actions during the live scenario and does such comparison for all steps in the sequence of actions of ground-truth. The result is represented as either 0 for mismatch, and $1$ for match leading to a list of $0$’s and $1$’s. Therefore, the sum of this list is divided by the total number of steps and constitutes $X$, i.e. the correctness of the order of actions.
To calculate $Y$, the VR training application simply checks whether individual actions were performed correctly. For example, if the action is to stop the machine, the trainee needs to select the correct button out of four possible options present during the scenario. Similarly, to the previous paragraph, the total number of correct actions is divided by the total number of actions in the ground-truth leading to the mean average of the correctness of the actions.
The $0.3$, and $0.2$ constants were chosen to highlight the bias towards the correctness of the order of actions due to the fact that the live scenario aims to assess more the order of actions performed by the trainee with the regard to the correction of actions. The $\sqrt{0.25XY}$ component is there to highlight the intertwined nature of the metrics as well as to stabilise the output within the range of $[0, 1]$.

The VR training system implements this scoring model through real-time tracking of user interactions. Each trainee's performance is logged and compared against the predefined correct sequence, providing immediate feedback on both task execution accuracy and proper ordering. This ensures that the training environment accurately simulates real-world procedural compliance and enables users to refine their decision-making processes through iterative attempts.




\section{User Testing \& Evaluation}
\label{Evaluation}

\subsection{User Feedback and Survey Results}  

\begin{figure}[htp]
\centering
  \subfigure[VR]{\includegraphics[scale=0.37]{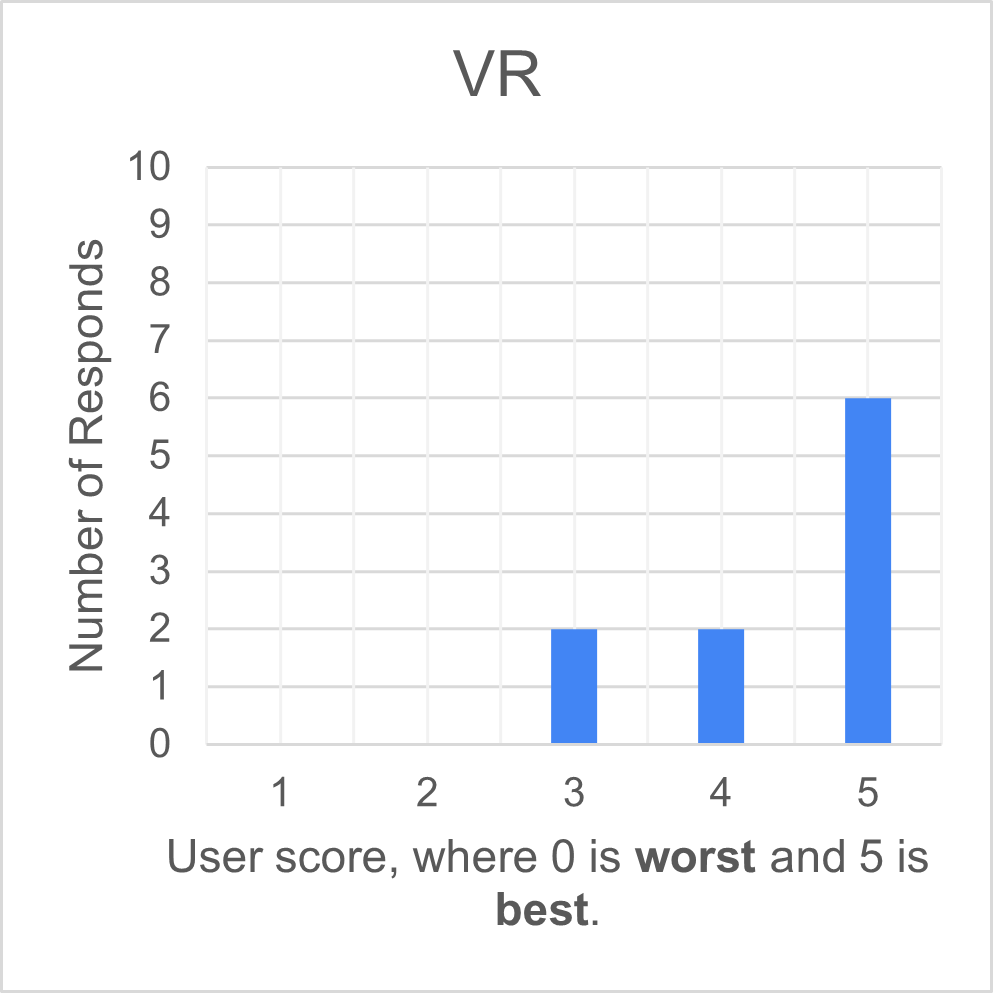}}
  \subfigure[Manual]{\includegraphics[scale=0.37]{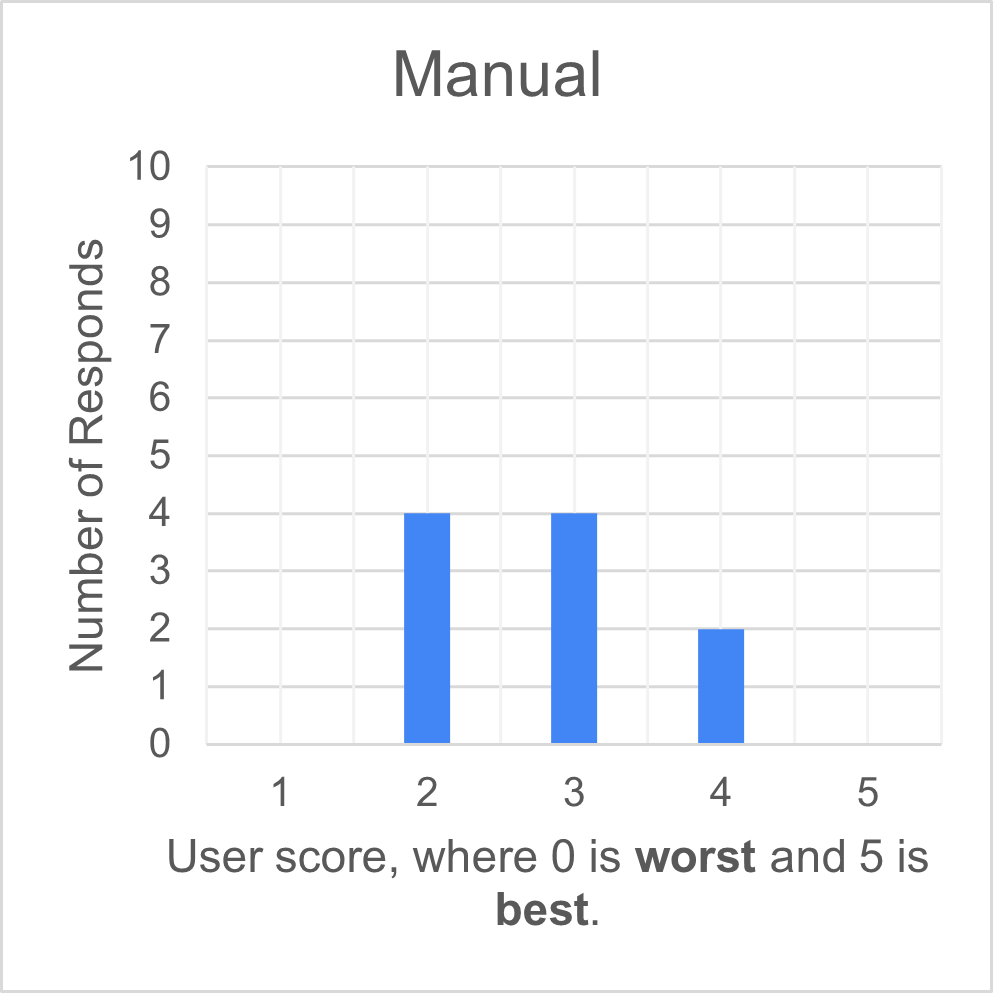}}
  \subfigure[Video]{\includegraphics[scale=0.37]{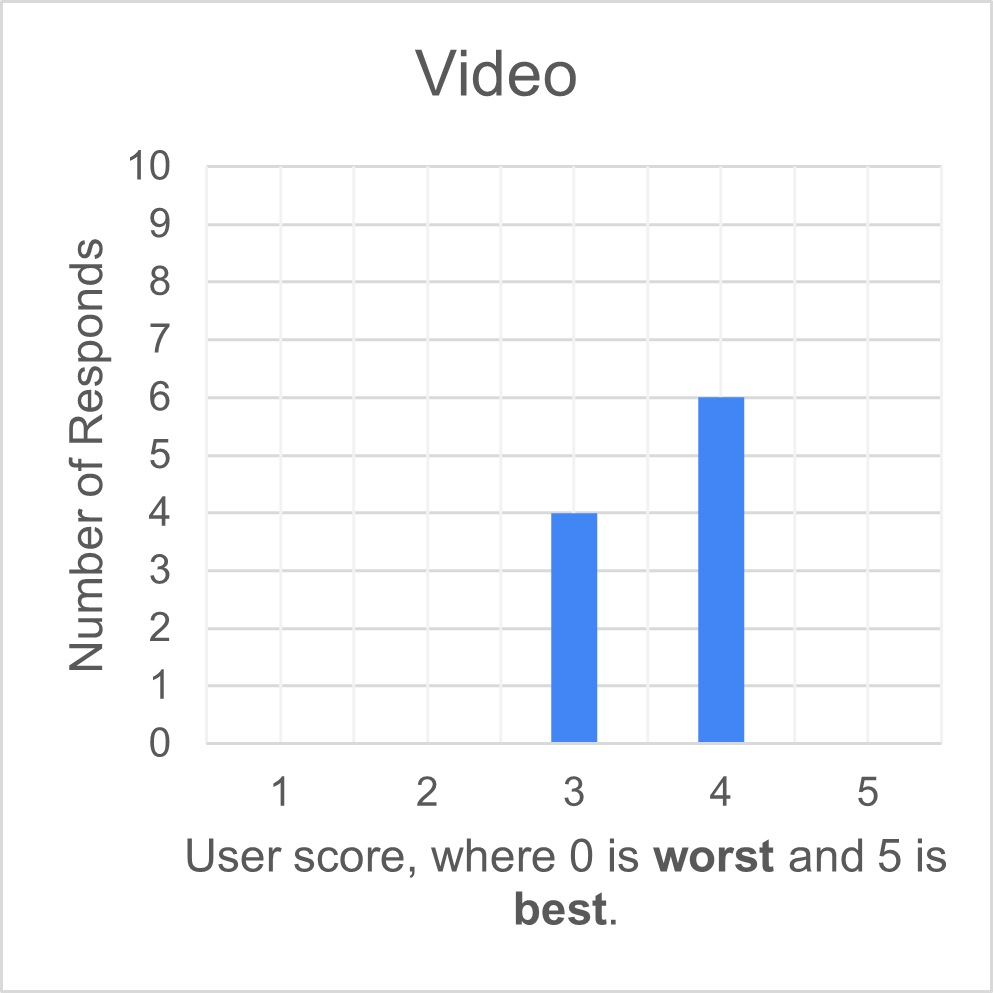}}
  \subfigure[VR]{\includegraphics[scale=0.37]{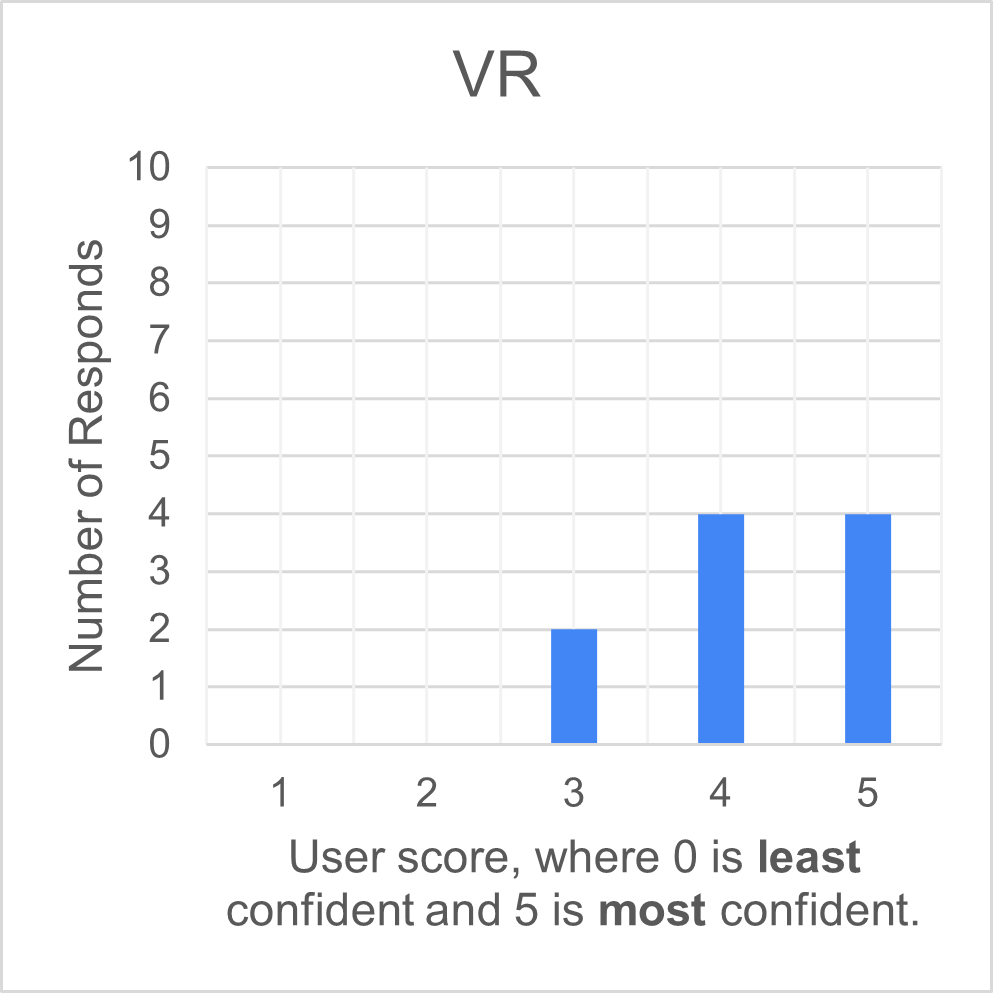}}
  \subfigure[Manual]{\includegraphics[scale=0.37]{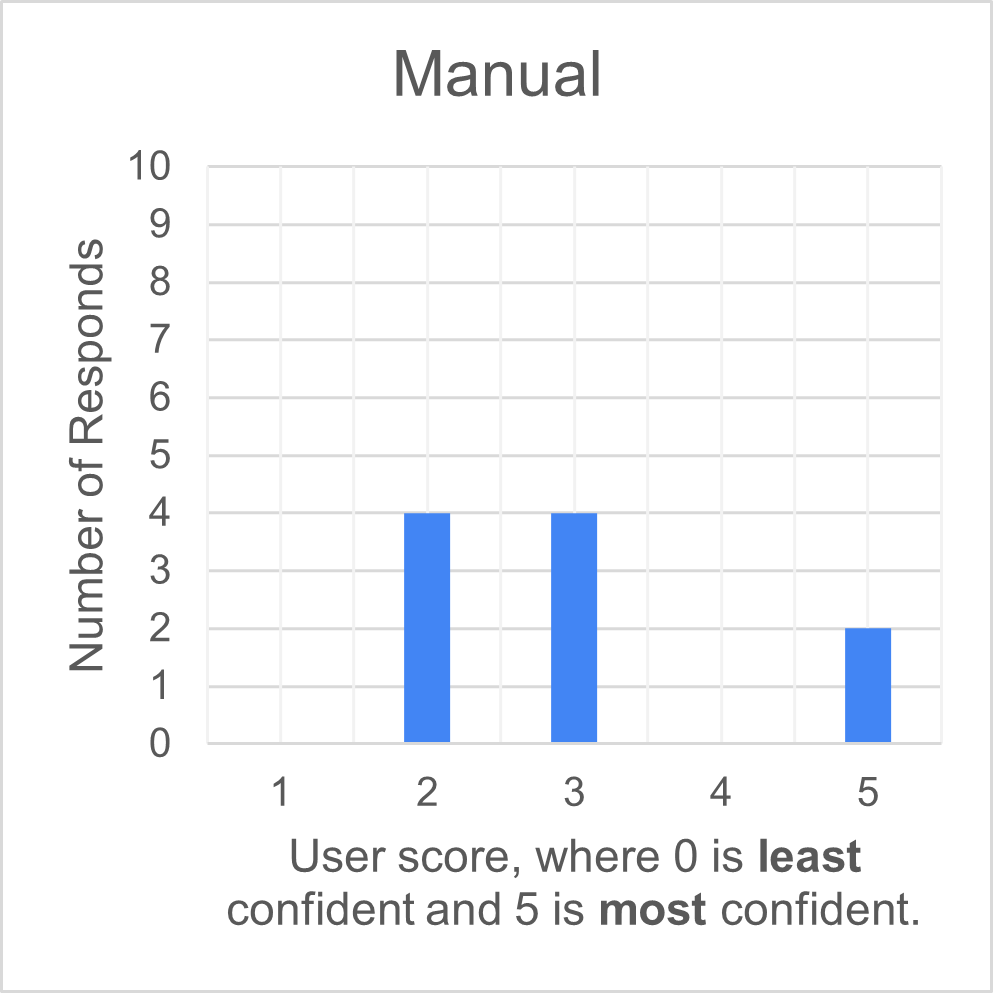}}
  \subfigure[Video]{\includegraphics[scale=0.37]{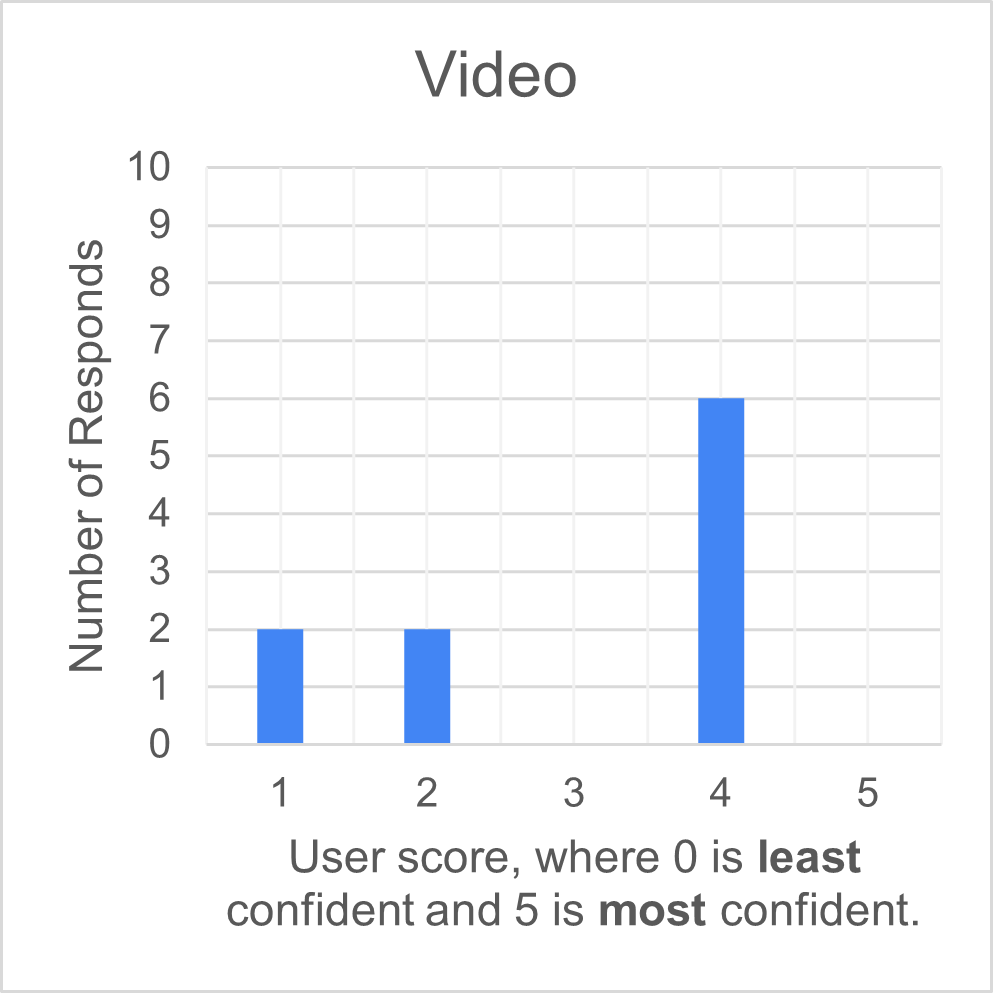}}
  \subfigure[VR]{\includegraphics[scale=0.37]{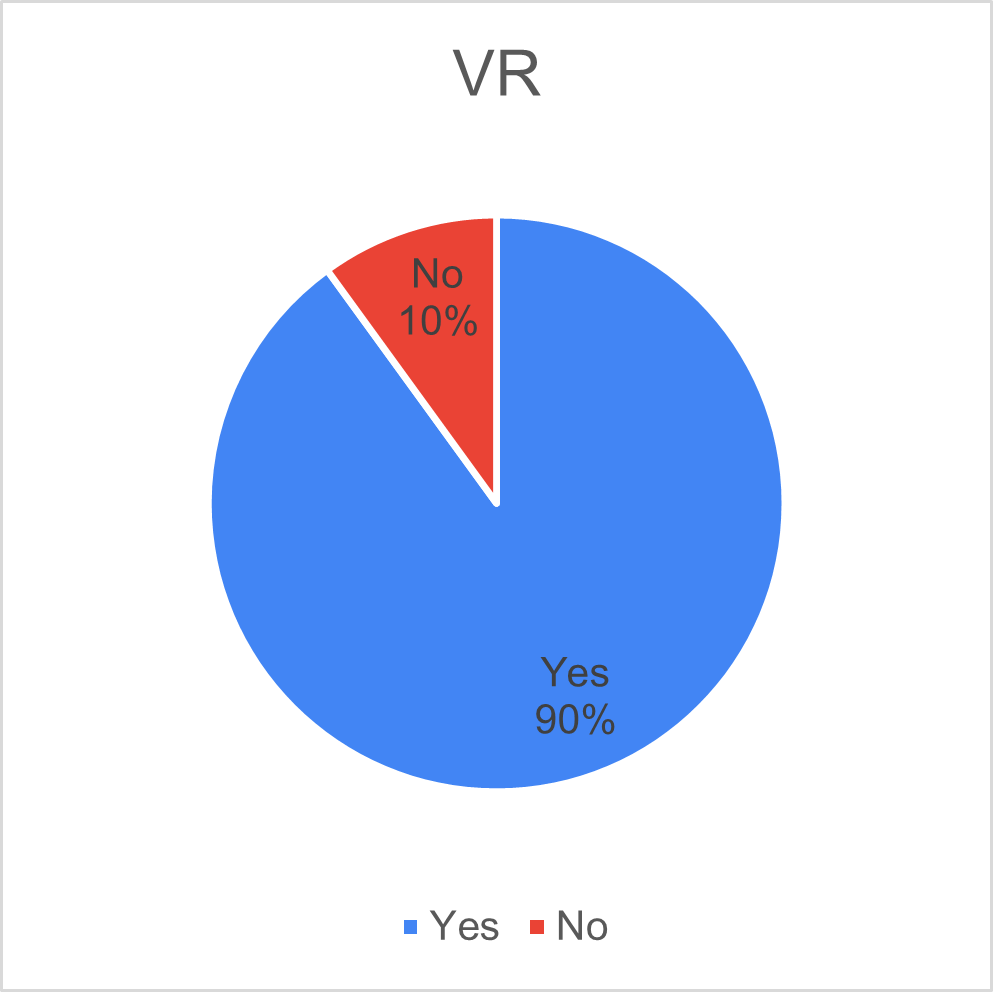}}
  \subfigure[Manual]{\includegraphics[scale=0.37]{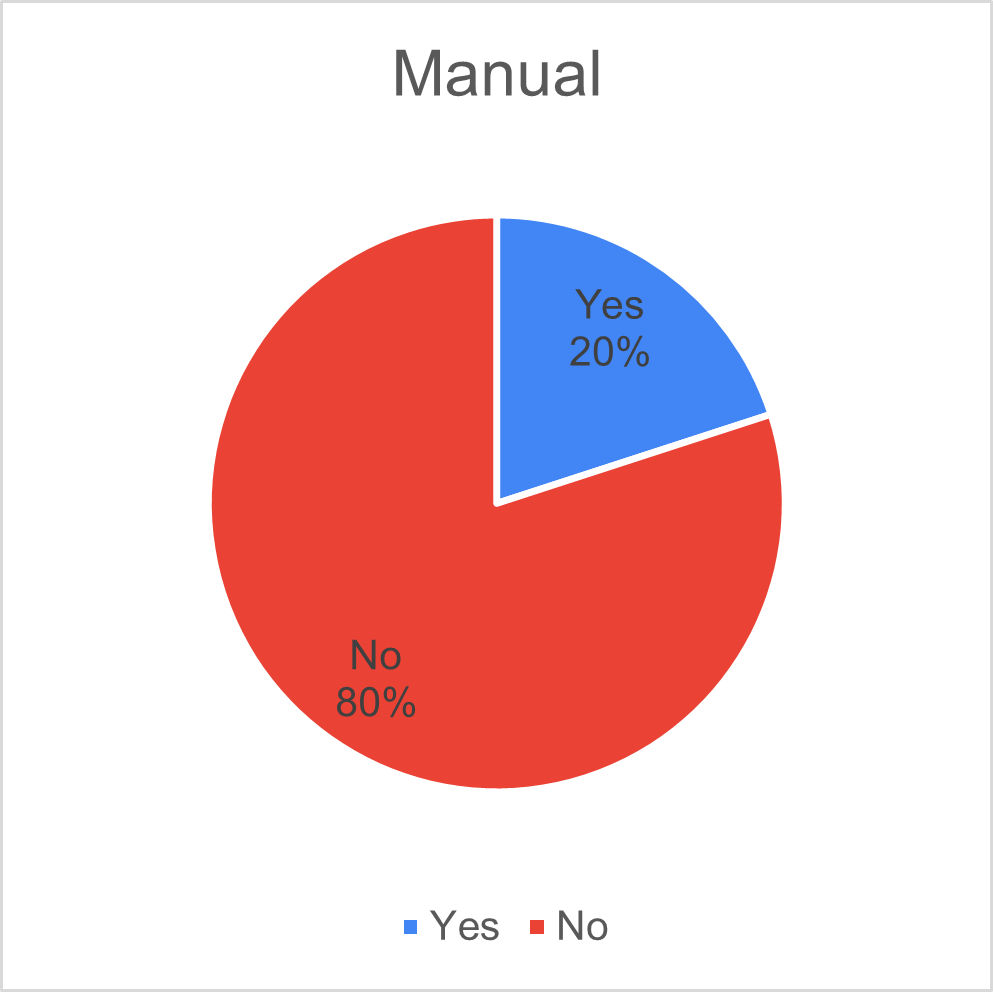}}
  \subfigure[Video]{\includegraphics[scale=0.37]{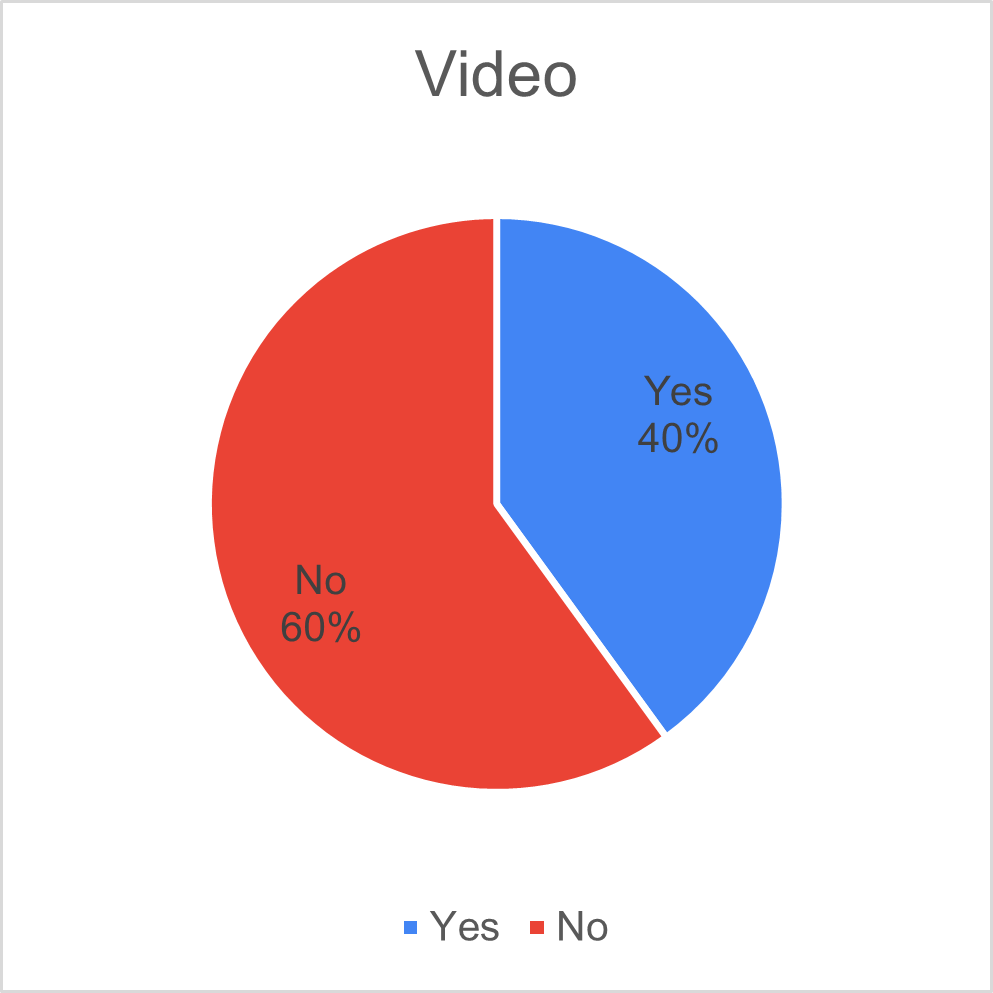}}
 \caption{User Satisfaction Comparison Examples: Q1-How well do you understand the health and safety procedures after completing the training?|Q2-How confident are you in applying the health and safety procedures in a real-world hazardous environment?|Q3-Do you believe the training method you used made it easier to stay focused throughout the learning process?}
 \label{fig:user_satisfaction_comparison} 
\end{figure}

The VR training application received predominantly positive feedback from participants, many of whom rated their experience with scores of 4 or 5 on a scale of 1 to 5. This high level of satisfaction highlights the overall success of the system in delivering an engaging and effective training experience. A standout feature for users was the intuitive navigation, which allowed them to easily explore the application without confusion or frustration. This ease of use was particularly appreciated by those who were new to virtual reality environments, as it minimized the learning curve and made the experience more accessible. Another key highlight was the live scenario element of the application, which was described as both engaging and highly relevant to real-world applications. Participants felt that this component provided a realistic and immersive way to practice health and safety procedures, reinforcing their understanding in a practical context. The training method was also praised for its ability to help users retain critical information, with many noting that it enhanced their confidence in applying health and safety measures in actual hazardous situations. Overall, user feedback underscores the system’s ability to effectively combine usability with practical training outcomes, positioning it as a valuable tool for health and safety education. Some examples of user satisfaction and feedback are given in Figure~\ref{fig:user_satisfaction_comparison}. In the provided samples it can be seen that users can learn more easily and confidently using the developed VR system than the technical manual or the video tutorials.

Moreover, Table~\ref{tab:synthetic-results} shows the summary statistics for user evaluations of the simulation, including confidence in application, retention effectiveness, engagement level, realism, and emergency preparedness. The results indicate that while users generally find the simulation effective for retention (Mean = 3.53) and moderately engaging (Mean = 3.13), there is notable variability in responses. Confidence in application is relatively strong (Mean = 3.40), but emergency preparedness receives the lowest rating (Mean = 2.80), suggesting a potential area for improvement.

\begin{table}[]
 \centering
 \caption{UX Feedback Evaluation}
 \label{tab:synthetic-results}
 \resizebox{1\columnwidth}{!}{
    \begin{tabular}{p{0.7cm}|p{1.5cm}|p{1.5cm}|p{1.6cm}|p{1.3cm}|p{1.7cm}|}
\hline
\textbf{}      & \textbf{Confidence in Application} & \textbf{Retention Effectiveness} & \textbf{Engagement Level} & \textbf{Realism of Simulation} & \textbf{Emergency Preparedness} \\ \hline
\textbf{count} & 15.00                              & 15.00                            & 15.00                     & 15.00                          & 15.00                           \\
\textbf{mean}  & 3.40                               & 3.53                             & 3.13                      & 3.13                           & 2.80                            \\
\textbf{std}   & 1.24                               & 0.99                             & 1.51                      & 1.25                           & 1.37                            \\
\textbf{min}   & 1.00                               & 2.00                             & 1.00                      & 1.00                           & 1.00                            \\
\textbf{25\%}  & 2.50                               & 3.00                             & 2.00                      & 2.00                           & 2.00                            \\
\textbf{50\%}  & 4.00                               & 3.00                             & 3.00                      & 3.00                           & 3.00                            \\
\textbf{75\%}  & 4.00                               & 4.00                             & 4.50                      & 4.00                           & 3.00                            \\
\textbf{max}   & 5.00                               & 5.00                             & 5.00                      & 5.00                           & 5.00                            \\ \hline
\end{tabular}
 }
\end{table}

\begin{table*}[]
 \centering
 \caption{Task Performance Summary}
 \label{tab:task_performance}
 \resizebox{1\textwidth}{!}{
\begin{tabular}{|l|c|c|c|p{6cm}|}
\hline
\textbf{Subtask}      & \textbf{Avg. Completion Time} & \textbf{Range}        & \textbf{Success Rate} & \textbf{Comments}                                                                                                                                 \\ \hline
MCQ Questions         & 20 sec                         & 15-30 sec               & $>90$\%                 & Users completed the MCQ Questions quickly and efficiently, indicating a well-designed and straightforward task.                                   \\ \hline
Interactive Questions & 3 min 7 sec                   & 33 sec - 3 min 24 sec & $>90$\%                 & Users required more time due to the interactive nature of questions but achieved full success. Some interactions could be refined.                \\ \hline
Live Scenario         & 2 min 24 sec                  & 55 sec - 3 min 35 sec & $>75$\%                  & This subtask had the widest range in completion times, suggesting some users struggled with aspects of the scenario. More guidance may be needed. \\ \hline
\end{tabular}
 }
\end{table*}

\begin{table}[]
 \centering
 \caption{Sample of Trainee Learning Performance Metrics}
 \label{tab:learning_metrics}
 \resizebox{1\columnwidth}{!}{
\begin{tabular}{|l|c|c|l|}
\hline
\textbf{SubTask}  & \textbf{MCQ}      & \textbf{Interactive} & \textbf{LiveScenario} \\ \hline
VRTSS mean        & \textbf{0.7000} & \textbf{0.5333}    & \textbf{0.4323}     \\
VRTSS std         & \textbf{0.3333} & \textbf{0.3583}    & \textbf{0.1098}     \\
VRTSS P-Value     & \textbf{0.0451} & \textbf{0.0441}    & \textbf{0.0001}     \\ \hline
Success Rate (\%) & \textbf{80.00\%}  & \textbf{60.00\%}     & \textbf{40.00\%}      \\ \hline
\end{tabular}
 }
\end{table}

The metrics indicate a generally positive reception of the training system across multiple dimensions. Confidence in application and retention effectiveness received some of the highest ratings, suggesting that users found the training both practical and memorable. Engagement level and realism of the simulation were also rated positively, though there is room for improvement in these areas to enhance immersion and user interest. Emergency preparedness received the lowest average rating, pointing to a potential need for more detailed scenarios or additional guidance to better equip users for high-stress situations. These insights provide valuable direction for refining the system to address specific user needs and expectations.

\subsection{Subtask Analysis}
\label{Experiment_Results}
The MCQs module proved to be an effective component of the training system, with $>90$\% success rate and an average completion time of just 20 seconds. This efficiency demonstrates that the questions were well-designed and straightforward, allowing participants to quickly assess their knowledge without confusion. Many users commented on the clarity of the questions and their relevance to the training objectives, which contributed to the high success rate. Furthermore, as illustrated in Table~\ref{tab:learning_metrics}, the majority of participants did not encounter significant difficulties with the multiple-choice question component of the training. This outcome was expected, as this section closely resembles conventional theory-centered training methods, such as standardized multiple-choice testing. The familiarity of this format enabled participants without prior exposure to the VR training application to quickly comprehend the training objectives and complete the task successfully. Nevertheless, there were instances in which participants were unable to complete the task due to a lack of familiarity with VR technology and the operational aspects of VR hardware, such as gesture recognition camera tracking. This finding suggests that an initial familiarisation phase with the VR hardware may be beneficial before commencing the training.

Similarly, the IQ module also achieved a $>90$\% success rate but required a longer average completion time of 3 minutes and 7 seconds. This module involved more complex tasks, such as interacting with virtual objects, which users generally found engaging despite occasional challenges. Some participants noted difficulties with object manipulation, suggesting that refining the interaction mechanics could enhance the user experience. In the interactive question section, participants were generally able to grasp the concept with ease and proceed through the training. In some cases, such as Participant 4 (as shown in Table~\ref{tab:learning_metrics}), performance in the IQ section was superior to that in the MCQ section, as it was perceived to be more intuitive and closely aligned with real-world scenarios. Overall, the performance amongst the participants remained the same as for MCQ.

Finally, the live scenario module, while highly engaging, had a lower success rate of $>75$\%. Participants spent an average of 2 minutes and 24 seconds completing this module, with significant variability in completion times. This variability highlights the need for clearer guidance or more intuitive design elements to help users navigate the scenarios effectively. Table~\ref{tab:task_performance} summarises the statistics for the trainees for each subtask. This final module of the training proved to be the most challenging for participants; however, it was also the most engaging. This module has been focused on the practical side of the training, i.e. learn-by-doing. For certain individuals, such as Participant 1, this section was comparatively easier, indicating that this training approach may be particularly beneficial for individuals who find traditional methods of training less effective.


\section{Conclusions}
\label{Conclusions}
Training employees in industrial operations can be a challenging task, especially for the high-pace  and high-risk environment of industrial manufacturing. The VR technology poses as a very promising solution for bridging the gap of on-hands experience while in a safe, controlled, and monitor environment that can gauge the interest of potential trainees. It also provides a platform for interactive and adaptive learning with better results. To this end, this study presents the development of a VR training system and methodology for industrial manufacturing. The methodology includes a novel metric, VRTSS Score, to measure the learning efficiency and capabilities of the trainies in a variety of tasks in three distinct subtasks, namely, i)Multiple-choice Questions, ii) Interactive Questions and iii) Live Scenarios, to gauge usability, knowledge retention, and user engagement. The experimental results highlight the system’s effectiveness, particularly in the MCQ and interactive question modules, which exhibited a $>90$\%  success rate and demonstrated clear user comprehension and efficiency. A comprehensive analysis of user feedback underscores the strengths of the VR training system, including its ease of navigation, immersive live scenario experiences, and high engagement levels. Despite these successes, we did find some limits, most notably in the Live Scenario subtask, where we had lower success rates and more completion time fluctuation, both of which pointed to possible usability issues. Based on these results, certain tweaks are in order, such making the interaction dynamics stronger, making the system guidance better, and simplifying the experience overall so it can cater to a wider range of user skills.

Future research will aim to improve task-specific interactions, optimize engagement-driven design components, and integrate additional user aid mechanisms to make this training platform more reliable and resilient. These updates will confirm the VR system's worth for more extensive uses in immersive training settings and further cement its place in health and safety education. The suggested virtual reality training system can develop into a powerful resource for training and education practices by systematically addressing these factors.


\section*{Match \& Contribution} 
This contribution aligns well with the theme of the ICE IEEE 2025 conference on "Autonomous and Self-organized Artificial Intelligent Orchestrator for a Greener Industry 4.0”. The paper is bridging the gap between theoretical instruction and practical competence remains a persistent challenge in industrial manufacturing training. Traditional methods often fall short in equipping workers with the necessary hands-on skills, especially in increasingly complex, technology-driven environments. This study proposes a Virtual Reality (VR)-based training framework designed to enhance learning outcomes through immersive, high-fidelity simulations and adaptive feedback mechanisms. The system features intuitive gesture-based controls to accommodate users of varying skill levels and introduces a novel performance metric—the VR Training Scenario Score (VRTSS)—for dynamic trainee evaluation. Experimental results demonstrate significant improvements in knowledge retention, task accuracy, and overall training effectiveness, underscoring VR's potential as a scalable and effective alternative to conventional training approaches.

\section*{Acknowledgement}
This work was funded by UK Research and Innovation (UKRI) under the UK government’s Horizon Europe funding guarantee [grant number 10047653] and funded by the European Union [under EC Horizon Europe grant agreement number 101070181 (TALON)].

\bibliographystyle{IEEEtran}
\bibliography{bib}

\end{document}